\documentclass[aps,prl,twocolumn,preprintnumbers,groupedaddress,superscriptaddress,floatfix,tightenlines,reprint]{revtex4-1}
\usepackage{mathrsfs,natbib}
\usepackage{graphics,epsfig,color, graphicx}
\usepackage{verbatim}

\usepackage{graphicx,epsf,amssymb,bbm,amsbsy,amsfonts,amssymb,amsmath}
\usepackage{hyperref}

\newcommand{\eq}{\begin{equation}}
\newcommand{\eqe}{\end{equation}}

\newcommand{\eqa}{\begin{eqnarray}}
\newcommand{\eqae}{\end{eqnarray}}

\hypersetup{
    colorlinks=true,       
    linkcolor=red,          
    citecolor=blue,        
    filecolor=magenta,      
    urlcolor=blue           
}

\begin{document}

\title{Real-Time Feynman Path Integral Realization of Instantons 
}
\preprint{FTPI-MINN-14/18, UMN-TH-3343/14}

\author{Aleksey Cherman}
\email{acherman@umn.edu}
\affiliation{Fine Theoretical Physics Institute, Department of Physics, University of Minnesota, Minnesota, MN 55455, USA}

\author{Mithat \"Unsal}
\email{unsal.mithat@gmail.com}
\affiliation{Department of Physics, North Carolina State University, Raleigh, NC, 27695}

\begin{abstract}
In Euclidean path integrals, quantum mechanical tunneling amplitudes are associated with instanton 
configurations.  We explain how tunneling amplitudes are encoded in real-time Feynman path integrals. 
The essential steps  are   borrowed from Picard-Lefschetz  theory 
 and resurgence theory. 
\end{abstract}


\maketitle

{\it Introduction}---The real-time Feynman path  integral\cite{RevModPhys.20.367,*feynman1965quantum}  is one of the several equivalent formulations of quantum mechanics (QM) and quantum field theory (QFT).  
 The amplitude of an event is found by summing over all path histories with the contribution of each path $\gamma$ weighed by its action $S[\gamma]$ via the factor $e^{i S [\gamma] /\hbar}$.
  For  real paths,  this is a sum over pure phases. The real-time path integral formulation is very useful in many settings, and yields beautiful explanations of many quantum phenomena, such as 
  the interference patterns seen in double-slit experiments and the Aharonov-Bohm effect.  
  Yet surprisingly,  some very basic quantum phenomena do not have a known description in the real-time path integral. For instance, tunneling phenomena,  which are 
  well-understood in the operator (Hamiltonian) formalism, as well as in the Euclidean-time path integral formulation,  are not yet understood in the real-time Feynman path integral formulation\footnote{For an earlier attempt at grappling with instantons in real time see {\it e.g.} \cite{Levkov:2004ij}. }. 
  
We focus on a particle in a symmetric double-well potential as a paradigmatic example.   The level splitting  
  between the ground state and first excited state is given by a non-perturbative factor, $\sim \exp[-A/(g^2 \hbar)]$ where $A$ is a pure number and $g$ is a parameter controlling the depth of the wells.  It  is a  consequence of  tunneling in the operator formalism  or instanton saddles\cite{Belavin:1975fg} in 
the Euclidean path integral formulation,  but its origin  is mysterious in the Feynman path  integral.

Understanding such non-perturbative phenomena directly in real-time path integrals is likely to be very important.  Euclidean path integrals give deep insights into the non-perturbative properties of QM and QFT,
but there is a huge range of problems for which they are not useful, such as calculations of scattering amplitudes or non-equilibrium observables.  These questions are more naturally handled using real-time path integrals.  Yet the current understanding of real-time path integrals beyond the perturbative level is extremely rudimentary, as is highlighted by the fact that even tunneling phenomena are not understood in the framework.    Our results are a step toward developing a \emph{non-perturbative} understanding of real-time path integrals.

{\it Euclidean-time instantons}---To set notation we review the standard path integral treatment of tunneling in QM. The    \emph{real-time} Feynman 
path integral representation of the amplitude $\mathcal{A}$ for a particle initially at position $x_i$ at time $t_i$ to end up at position $x_f$ at time $t_f$ is
\begin{align}
\mathcal{A}_{x_i, t_i}^{x_f,t_i} = \int_{x(t_i) = x_i}^{x(t_f) = x_f} d[x(t)] \, e^{\frac{i}{\hbar} S[x]}
\end{align}
where $S$ is the classical action.  For a non-relativistic bosonic particle moving in a symmetric double-well potential the action can be written as
\begin{align}
S[x] =  \frac{1}{2g^2}\int_{t_i}^{t_f} dt  \left[ (\partial_t x)^2 - (x^2-1)^2\right]
\end{align}
Our semiclassical treatment of tunneling will be justified if $g \ll 1$.  If the particle is initially near e.g. $x = -1$ at $t =t_i$ the amplitude for it to end up near $x=+1$ will be zero to any order in a perturbative expansion in $g$. 

\begin{figure*}[h!t]
  \centering
\includegraphics[width=0.95\textwidth]{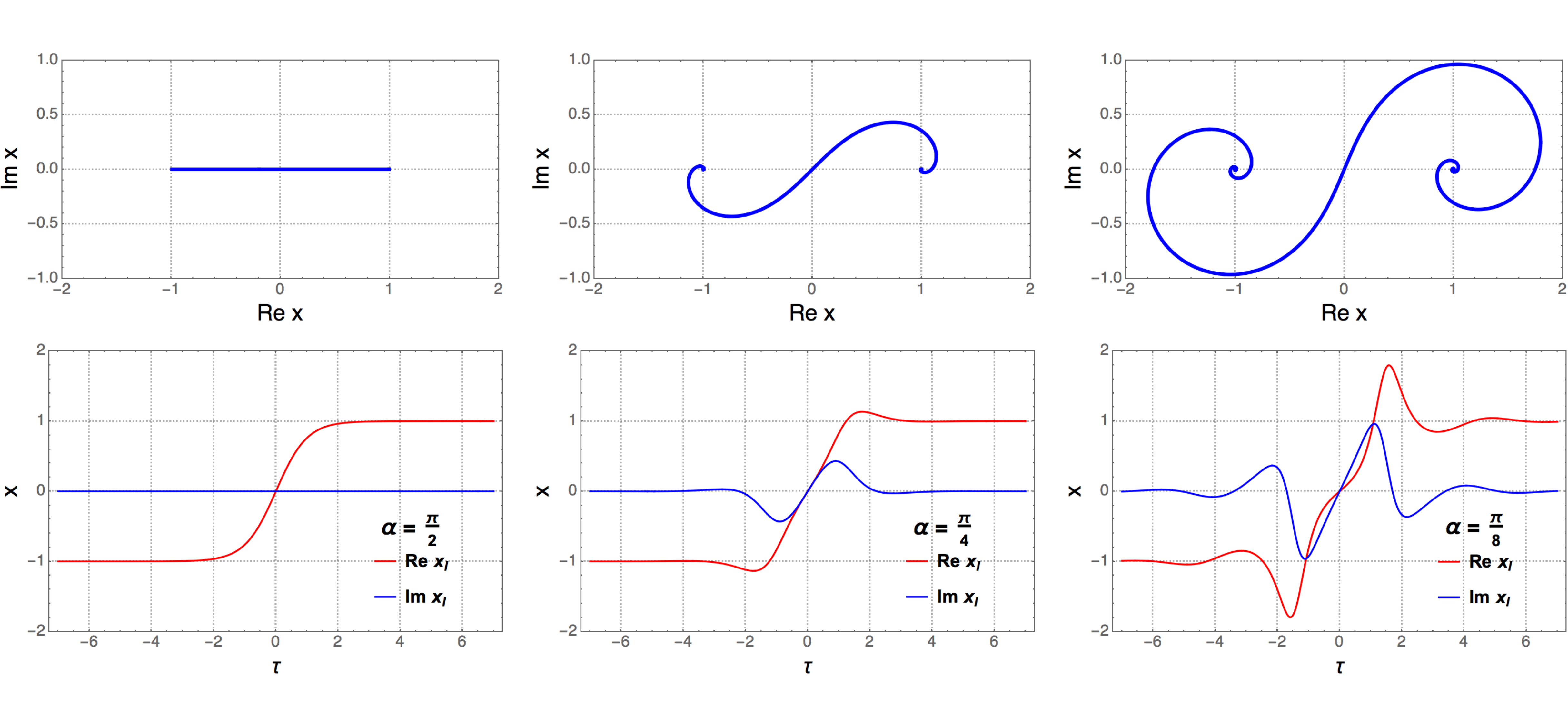}
  \caption{(Color Online.) The complexified instanton configuration \eqref{eq:ComplexInstanton}.  {\bf Left column:} $\alpha = \pi/2$ (Euclidean instanton). {\bf Center column:} $\alpha = \pi/4$.  {\bf Right column:} $\alpha = \pi/8$.  The top row shows the trajectories of \eqref{eq:ComplexInstanton} for $\tau \in (-\infty, +\infty)$, while the bottom row shows the real and imaginary parts of \eqref{eq:ComplexInstanton} as a function of $\tau$.  }
  \label{fig:lowAlpha}
\end{figure*}

Let us now focus on the tunneling amplitude from $x_i=-1$ at $t_i = -\infty$ to $x_f=+1$ at $t_f = + \infty$.  The standard way to compute this amplitude using path integrals involves moving to Euclidean time by a Wick rotation $t \to -i \tau, x(t) \to x(\tau)$.  The Euclidean action
\begin{align}
S_E = \frac{1}{2g^2}\int_{-\infty}^{\infty} d\tau  \left[ (\partial_\tau x)^2 + (x^2-1)^2\right].
\end{align}
is then positive semi-definite and enters the path integral as $e^{-S_E}$.  The Euclidean equation of motion
$\partial^2_\tau x(\tau) = 2 x(\tau) \left(x(\tau)^2-1\right)$
has two finite-action solutions, the instanton $x^{+}_{I}$ and anti-instanton $x^{-}_{I}$:
\begin{align}
x^{\pm}_{I}(\tau) = \pm \tanh (\tau - \tau_0) ,
\label{eq:EuclideanInstanton}
\end{align}
where $\tau_0$ is a zero mode parameter.  These solutions interpolate from $x = \mp 1$ to $x= \pm1$ as $\tau$ goes from $-\infty$ to $+\infty$.  In what follows we will focus on $x^{+}_I$ and will drop the superscript.  The Euclidean action of \eqref{eq:EuclideanInstanton}  is $S_{I} = \frac{4}{3 g^2}$, so that \eqref{eq:EuclideanInstanton} is a saddle-point of the Euclidean path integral, with contributions weighed by $e^{-\frac{4}{3g^2}}$.  For $g \ll 1$ the desired tunneling amplitude can be computed using a steepest-descent approximation, and is given by
\begin{align}
\mathcal{A} = \sqrt{\frac{32}{ \pi g^2}} e^{-\frac{4}{3g^2 \hbar}}
\label{eq:TunnelingAmplitude}
\end{align}
where the prefactor is obtained by integrating over Gaussian and zero mode fluctuations around $x = x_I(\tau)$.  

{\it Real-time instantons}---The Euclidean-time derivation is not fully satisfying, as physics happens in real time.  

\begin{figure*}[htbp]
  \centering{
\includegraphics[width=0.95\textwidth]{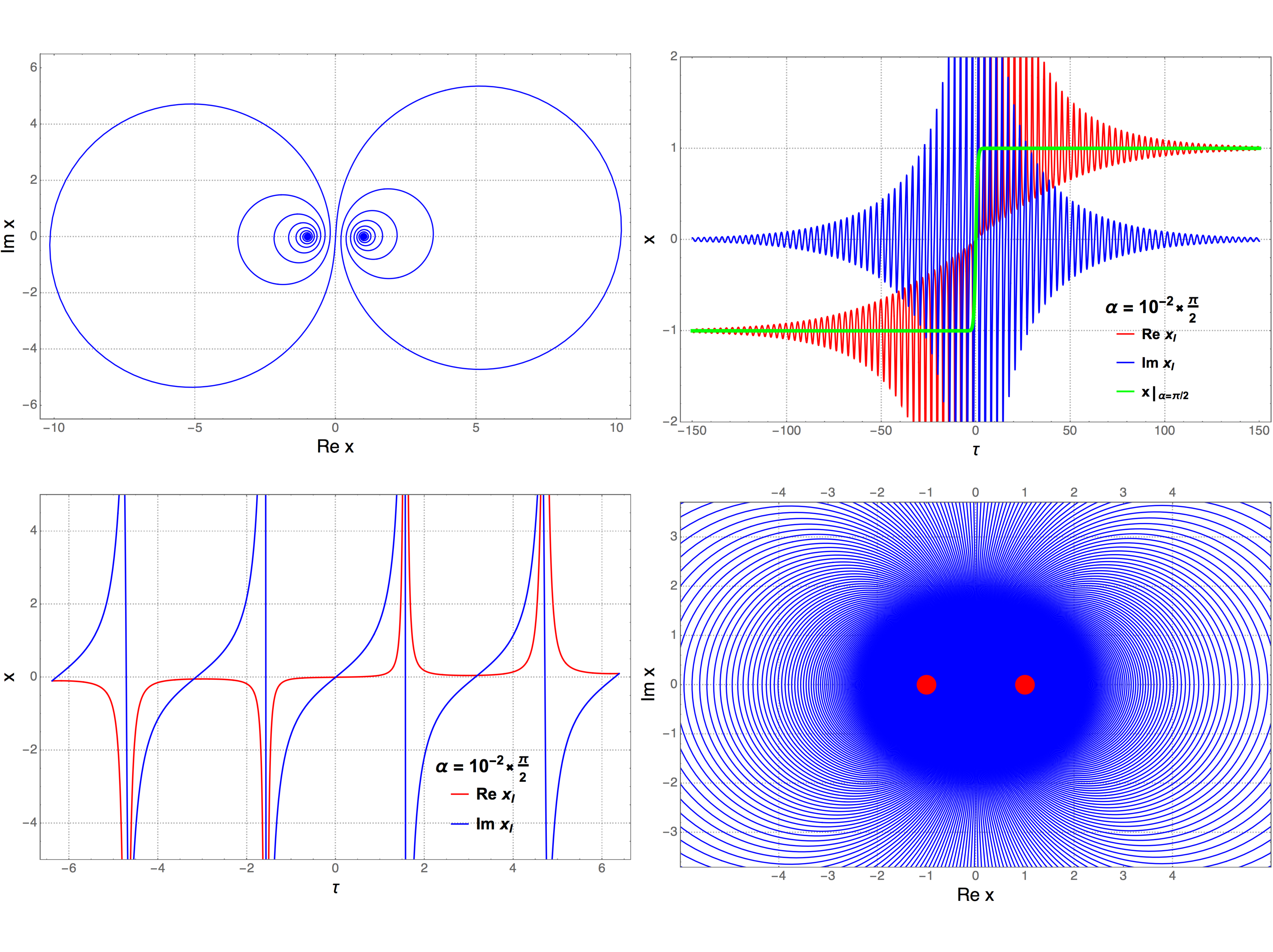}}
  \caption[justification=justified]{(Color Online.) {\bf Upper-left:}  trajectory of \eqref{eq:ComplexInstanton} for $\alpha = 4 \times 10^{-2} \times \frac{\pi}{2}$ in complex configuration space. {\bf Upper-right:} real and imaginary parts of  \eqref{eq:ComplexInstanton}  for $\alpha =  1 \times 10^{-2} \times \frac{\pi}{2}$, with a green $\alpha=\pi/2$ trajectory shown for comparison.  {\bf Lower-left:} illustration of the manner in which the imaginary part of \eqref{eq:ComplexInstanton} approaches the singular configuration \eqref{eq:SingularInstanton} for small $\alpha$ (here $\alpha = 1 \times 10^{-2} \times \frac{\pi}{2}$), while the real part   approaches to a distribution, 
  a signed Dirac comb.  {\bf Lower-right:} the trajectory for $\alpha =  1 \times 10^{-3} \times \frac{\pi}{2}$, with red dots marking the centers of the wells $x =  \pm 1$. In the real-time limit the real-time instanton appears to become a space-filling curve in complexified configuration space.  }
  \label{fig:highAlpha}
\end{figure*}

Yet there is a good reason that instantons were never formulated in the real-time Feynman path integral.    A natural attempt to do this involves looking for finite-action solutions of the real-time equation of motion
\begin{align}
-\partial^2_t x(t) = 2 x(t) \left(x(t)^2-1\right) .
\label{eq:RealTimeEoM}
\end{align}
Since we are looking for an answer of the form $e^{i S/\hbar} = e^{-\frac{A}{g^2\hbar}}$ the action of the solution would have to be imaginary.  That means that a `real-time instanton' configuration has to be complex.  Such a notion raises  major conceptual issues:    \\
{\bf (A)}
If the field space is complexified,  how do we avoid an undesired doubling of degrees of freedom in the complexified path integral?    \\
{\bf (B)} 
Relatedly, how do we avoid including pathological directions in the complex field space which may yield exponentially large contributions?  Over what sub-space of fields (complex paths) should one  integrate over?

 Setting these important questions aside for the moment, we note that it is tempting to guess that the desired saddle field configurations might be given by the Wick rotation of the Euclidean instanton back to real time.    Indeed, \eqref{eq:RealTimeEoM}  does have the formal solution
\begin{align}
x(t) = i \tan (t-t_0)
\label{eq:SingularInstanton}
\end{align}
which is a Wick rotation of \eqref{eq:EuclideanInstanton}.  But \eqref{eq:SingularInstanton} looks non-sensical, for many reasons:  \\
{\it (i)} 
It is pure-imaginary-valued, so it cannot possibly give a contribution like $e^{i S/\hbar} = e^{-\frac{A}{g^2\hbar}}$, and  cannot describe a path interpolating from $x = - 1$ to $+1$.  \\
{\it (ii) } It does not appear to have anything to do with tunneling from $x = - 1$ to $+1$, since it is never localized near the bottoms of the wells. \\
 {\it (iii) } 
It has singularities at $t = (2n+1)\pi/2, n \in \mathbb{Z}$, and these singularities are not integrable, so that the action is not well-defined.

Until fairly recently this is where an analysis would have had to stop.  Fortunately, there has recently been major progress in the understanding the complexification of functional integrals \cite{Pham1983,Witten:2010cx,*Witten:2010zr,Tanizaki:2014xba} and the structure of the semiclassical expansion of path integrals in QM \cite{Bogomolny:1980ur,*ZinnJustin:1981dx,*Dunne:2014bca} and QFT \cite{Argyres:2012vv,*Argyres:2012ka,*Dunne:2012ae,*Dunne:2012zk,*Cherman:2013yfa,*Basar:2013eka,*Cherman:2014ofa,*Misumi:2014jua,*Misumi:2014raa} via resurgence theory\footnote{There are also related developments in string theory, see {\it e.g.} \cite{Marino:2007te,*Marino:2008ya,*Marino:2008vx,*Aniceto:2011nu,*Schiappa:2013opa}.}.   
Motivated by these developments, let us see what happens if we do not insist on working in either  purely real or imaginary time. 

Specifically, let us start in real time $t$ and do a \emph{complex} Wick rotation $t \to \tau e^{ - i \alpha}, \phi(t) \to \phi(\tau)$.  In contrast to purely real or purely imaginary time, now the action itself becomes complex:
\begin{align}
S_{\alpha} = \frac{e^{-i\alpha}}{2g^2} \int_{-\infty}^{+\infty} d\tau \, \left[ e^{+2i\alpha} (\partial_\tau x)^2 - (x^2-1)^2\right] .
\end{align}
Hence issues (A) and (B) with the complexification of configuration space in path integrals must be deal with, and we do so below.  For small $\alpha$, $\tau \approx t$ and $S_{\alpha} \approx S$, while for $\alpha = \pi/2$ $\tau$ becomes the standard Euclidean time coordinate and $iS_{\pi/2} = -S_E$.   It turns out that an infinitesimal $\alpha$ acts as a crucial regulator which is necessary to make sense of real-time instantons.  

The $\alpha$-dependent equation of motion
$-e^{2i\alpha} \partial^2_\tau x(\tau) = 2 x(\tau) \left(x(\tau)^2-1\right) $ has the solutions
\begin{align}
x_{I}(\tau) = \pm \tanh\left[(\tau - \tau_0) e^{ - i(\alpha - \pi/2)} \right]
\label{eq:ComplexInstanton}
\end{align} 
To make \eqref{eq:ComplexInstanton} a legitimate solution one must complexify the space of path  histories  from $x: \mathbb{R} \to \mathbb{R}$ to  $x: \mathbb{R} e^{-i \alpha} \to \mathbb{C}$.   While \eqref{eq:ComplexInstanton} looks like a simple analytic continuation of \eqref{eq:EuclideanInstanton}, the conceptual issues raised by allowing such an continuation are quite complex.

Naively, precisely at $\alpha = 0$ \eqref{eq:ComplexInstanton} reduces to \eqref{eq:SingularInstanton}, although we are about to see that the limit is subtle.  Consider the situation for $0 < \alpha \le \pi/2$~\footnote{The instanton and anti-instanton get interchanged if $-\pi/2 \le  \alpha < 0$.}.   Remarkably, the imaginary parts of the complex instanton solution \eqref{eq:ComplexInstanton} vanish for large $|\tau|$, and  $x_{I}$ approaches $\mp 1$ as $\tau \to \mp \infty$!   The behavior of the complexified instanton at generic $\alpha$ values is summarized in Fig.~\ref{fig:lowAlpha} and  Fig.~\ref{fig:highAlpha}.
For generic $0< \alpha \le \pi/2$ field starts at $x = -1$ at $\tau = -\infty$, whips around in the left half of complex-$x$ plane for $\tau < \tau_0$, then crosses to the right half of the plane at $\tau = \tau_0$, whips around $x=+1$ for $\tau > \tau_0$, and then settles down at $x=+1$ at $\tau = +\infty$.

 The instanton remains  smooth for any $\alpha \neq 0$, but the behavior in the real time limit $\alpha \to 0$ is singular, 
  as illustrated in Fig.~\ref{fig:highAlpha}.   The $\mathrm{Re}\, x <0$ and  
$\mathrm{Re}\, x>0$ half-planes form two basins of attraction for the repulsive and attractive fixed points $x=-1$ and $x=+1$ respectively.  After starting at $x = -1$ at $\tau  = - \infty$, as $\alpha \to 0$ the instanton spends more and more time whipping around in the $\mathrm{Re}\, x <0$ region in a huge number of more and more tightly spaced increasingly large arcs for $\tau < \tau_0$, then jumps over to the $\mathrm{Re}\, x >0$ region and winds closer and closer to $x=+1$ in the same manner, until settling at $x = +1$ at $\tau = +\infty$.   From the bottom-right plot in Fig.~\ref{fig:highAlpha}, the complex instanton trajectory appears to become a space-filling curve in the real-time limit.   

In the strict $\alpha=0$ limit,  the  imaginary part of the instanton solution  is given by \eqref{eq:SingularInstanton}, while the real part approaches to a distribution,  given by 
\begin{align}
\mathrm{Re}\, x = \pi {\rm sign}( t -t_0)  \sum_{k=-\infty}^{\infty} \delta ( t -t_0  -  \frac{\pi }{2}  (2k-1) ) 
\end{align}
which is a signed Dirac comb.  The time average of $ \langle \mathrm{Re}\, x \rangle_T   $ and $\langle \mathrm{Im}\, x \rangle_T$   (which we define as 
$\langle \cdot \rangle_T=   \frac{1}{T} \int_{\tau_0}^{T}  \cdot \; d\tau $
 with $T = \pi k +\tau_0, k \in \mathbb Z$)
for any $\alpha$,
including $\alpha=0$,
essentially follows the pattern of a  Euclidean instanton, 
\begin{align}
&\langle \mathrm{Re}\, x \rangle_T  = \left\{ \begin{array}{ll}   +1,  & \qquad \tau > \tau_0 \\ 
-1,  & \qquad \tau < \tau_0 
\end{array} \right. \cr
&\langle \mathrm{Im}\, x \rangle_T  = 0 
\end{align}
as illustrated by Fig.~\ref{fig:highAlpha}. 

Next, we show that the action $-i S_{\alpha}$ of the instanton is always precisely $4/(3g^2)$ for any $\alpha$, with the action at $\alpha = 0$ defined by continuation from $|\alpha| > 0$.
The Lagrangian evaluated on \eqref{eq:ComplexInstanton} oscillates wildly through the complex plane as $\alpha \to 0$.  Nevertheless, $-i S_{\alpha}[x_{I}]$ turns out to be \emph{independent} of $\alpha$. 
To show this we rewrite
\begin{align}
S_\alpha(x) = \frac{e^{-i\alpha}}{2g^2} \int_{-\infty}^{+\infty} d\tau\left( \left[e^{+i\alpha} \dot{x}  - W'(x) \right]^2 + 2 e^{+i\alpha}  \dot{x} W'\right) \nonumber
\end{align}
where $' \equiv \partial_x, \dot{}  \equiv \partial_{\tau}$, and $(W')^2 = V(x)$.  But  \eqref{eq:ComplexInstanton} satisfies the first-order equation $e^{+i\alpha}  \dot{x} - W'(x) = 0$, so
\begin{align}
-i S(x_I) = \frac{1}{g^2} \int_{-\infty}^{+\infty} d\tau \dot{x}_I  W'(x_I)= \frac{4}{3g^2}
\end{align}
We conjecture  that the solution with minimal non-zero action is \eqref{eq:ComplexInstanton} for any $\alpha$.  This is obvious at $\alpha = \pi/2$, while at $\alpha = 0$, resurgence theory implies that the minimum-action saddle must have action  $-iS(x_I)$ due to the large-order structure of perturbation theory.   We leave a detailed exploration of this conjecture to future work.

The question of whether singular but finite-action field configurations contribute to path integrals is controversial\cite{Harlow:2011ny}.  Our results support the notion that such configurations do contribute, since \eqref{eq:ComplexInstanton} is a finite action configuration whichis associated with a dramatic physical effect of level splitting.  Hence it is physically clear that it \emph{must} be included it  in the evaluation of the path integral, even though it becomes singular and discontinuous in the strict $\alpha=0$ limit.

{\it Lefshetz thimbles}---To find the contribution of the complex instantons to observables in the real time limit, it is necessary to take into account fluctuations around the instanton saddle points.  As we already noted this is a subtle problem due to the complexification of the field space, and involves understanding the analytic continuation of path integrals.  A program to study this issue was recently initiated in \cite{Witten:2010cx,*Witten:2010zr}.

A sensible complexification of the integration cycle of an integral which initially has a real domain of integration must not double the number of degrees of freedom.  A full understanding of the analytic continuation of the integral thus involves finding a basis of convergent middle-dimensional integration cycles in the complexified integral, and then working out how the original real cycle decomposes into the complex cycles.   Ref.~\cite{Witten:2010cx,*Witten:2010zr} showed that 
 in favorable circumstances, this can be done by using Picard-Lefshetz theory\footnote{How to carry out this program completely for configuration-space path integrals is an open problem, but the present understanding is already sufficient for our Gaussian analysis.}.  Every finite-action saddle point of the path integral is associated to a \emph{convergent} integration cycle called a `Lefshetz thimble' --- a multidimensional generalization of a steepest-descent path.  The thimbles are solutions of the downward flow equations 
  for the Morse function $\mathrm{Re}\,[i S]$.   For us, the flow equation is 
\begin{align}
\frac{\partial x}{\partial s} = -\frac{e^{+i(\alpha-\pi/2)}}{\overline{g^2}}\left[ e^{-2i\alpha} \partial^2_{\tau}\bar{x} +  2 \bar{x} \left(\bar{x}^2-1\right) \right]
\label{eq:FlowEquation}
\end{align} 
where  $x$ is  complex and $s$ is the flow `time' along the thimble.  The fluctuations around $x_I$ are described by the solutions of \eqref{eq:FlowEquation} with the initial condition $x(s \to -\infty, \tau) = x_I(\tau)$.  The structure of the Gaussian fluctuations can be determined by writing $x(s,\tau) = x_I(\tau) + \delta x(s,\tau)$ with $\delta x(s\to-\infty,\tau)=0$ and expanding \eqref{eq:FlowEquation} to $\mathcal{O} (\delta x)$:
\begin{align}
\frac{\partial \delta x}{\partial s} = -\frac{e^{+i(\alpha-\frac{\pi}{2})}}{g^2} \left[ e^{-2i\alpha} \partial^2_{\tau} + 6 \overline{x}^2_I(\tau)-2 \right] \overline{\delta x}
\label{eq:GaussianFlowEquation}
\end{align} 
The existence of the flow equations means that it is possible to choose sensible integration cycles in the complexified path integral, and in the Gaussian-fluctuation limit the determination of our desired cycle reduces to finding the solution of \eqref{eq:GaussianFlowEquation}.  
  By moving to the variable $\tau' = \tau \exp[-i(\alpha - \pi/2)]$ and rescaling $\delta x$ one can map \eqref{eq:GaussianFlowEquation} to the \emph{Euclidean} flow equation.  This implies that integrating over the thimble amounts to calculating the standard fluctuation determinant (see \cite{Tanizaki:2014xba} for a nice discussion of this) multiplying $e^{-4/3g^2}$, and reproduces \eqref{eq:TunnelingAmplitude}.   

{\it Conclusions}---We have shown how to compute tunneling amplitudes between low-lying states, a hallmark non-perturbative feature of quantum mechanics, directly from a real-time Feynman path integral.  The amplitude is given by a real-time instanton saddle point, which requires a complexification of the space of paths and the addition an infinitesimal imaginary part to the time coordinate as a regulator for its definition as a smooth configuration.  In the real-time limit the instanton trajectory takes a wild ride  through complex configuration space on its way between the minima of the potential, with the path resembling a space-filling curve.  Yet its action remains equal to that of the Euclidean instanton.    Our results are a step toward developing a non-perturbative understanding of physics using the real-time Feynman path integral formalism.  

{\it Acknowledgements:}  We are very grateful to Y.~Tanizaki, G.~Ba\c{s}ar, D.~Dorigoni,  G.~Dunne, A.~Joseph, and M.~Shifman for discussions.   A.~C. acknowledges the partial support of the US DOE under grant number DE-SC0011842.

\bibliographystyle{apsrev4-1}
\bibliography{instantons,pcm}

\begin{thebibliography}{30}%
\makeatletter
\providecommand \@ifxundefined [1]{%
 \@ifx{#1\undefined}
}%
\providecommand \@ifnum [1]{%
 \ifnum #1\expandafter \@firstoftwo
 \else \expandafter \@secondoftwo
 \fi
}%
\providecommand \@ifx [1]{%
 \ifx #1\expandafter \@firstoftwo
 \else \expandafter \@secondoftwo
 \fi
}%
\providecommand \natexlab [1]{#1}%
\providecommand \enquote  [1]{``#1''}%
\providecommand \bibnamefont  [1]{#1}%
\providecommand \bibfnamefont [1]{#1}%
\providecommand \citenamefont [1]{#1}%
\providecommand \href@noop [0]{\@secondoftwo}%
\providecommand \href [0]{\begingroup \@sanitize@url \@href}%
\providecommand \@href[1]{\@@startlink{#1}\@@href}%
\providecommand \@@href[1]{\endgroup#1\@@endlink}%
\providecommand \@sanitize@url [0]{\catcode `\\12\catcode `\$12\catcode
  `\&12\catcode `\#12\catcode `\^12\catcode `\_12\catcode `\%12\relax}%
\providecommand \@@startlink[1]{}%
\providecommand \@@endlink[0]{}%
\providecommand \url  [0]{\begingroup\@sanitize@url \@url }%
\providecommand \@url [1]{\endgroup\@href {#1}{\urlprefix }}%
\providecommand \urlprefix  [0]{URL }%
\providecommand \Eprint [0]{\href }%
\providecommand \doibase [0]{http://dx.doi.org/}%
\providecommand \selectlanguage [0]{\@gobble}%
\providecommand \bibinfo  [0]{\@secondoftwo}%
\providecommand \bibfield  [0]{\@secondoftwo}%
\providecommand \translation [1]{[#1]}%
\providecommand \BibitemOpen [0]{}%
\providecommand \bibitemStop [0]{}%
\providecommand \bibitemNoStop [0]{.\EOS\space}%
\providecommand \EOS [0]{\spacefactor3000\relax}%
\providecommand \BibitemShut  [1]{\csname bibitem#1\endcsname}%
\let\auto@bib@innerbib\@empty
\bibitem [{\citenamefont {Feynman}(1948)}]{RevModPhys.20.367}%
  \BibitemOpen
  \bibfield  {author} {\bibinfo {author} {\bibfnamefont {R.~P.}\ \bibnamefont
  {Feynman}},\ }\href {\doibase 10.1103/RevModPhys.20.367} {\bibfield
  {journal} {\bibinfo  {journal} {Rev. Mod. Phys.}\ }\textbf {\bibinfo {volume}
  {20}},\ \bibinfo {pages} {367} (\bibinfo {year} {1948})}\BibitemShut
  {NoStop}%
\bibitem [{\citenamefont {Feynman}\ and\ \citenamefont
  {Hibbs}(1965)}]{feynman1965quantum}%
  \BibitemOpen
  \bibfield  {author} {\bibinfo {author} {\bibfnamefont {R.}~\bibnamefont
  {Feynman}}\ and\ \bibinfo {author} {\bibfnamefont {A.}~\bibnamefont
  {Hibbs}},\ }\href {http://books.google.com/books?id=14ApAQAAMAAJ} {\emph
  {\bibinfo {title} {Quantum mechanics and path integrals}}},\ International
  series in pure and applied physics\ (\bibinfo  {publisher} {McGraw-Hill},\
  \bibinfo {year} {1965})\BibitemShut {NoStop}%
\bibitem [{Note1()}]{Note1}%
  \BibitemOpen
  \bibinfo {note} {For an earlier attempt at grappling with instantons in real
  time see {\protect \it e.g.} \cite {Levkov:2004ij}.}\BibitemShut {Stop}%
\bibitem [{\citenamefont {Belavin}\ \emph {et~al.}(1975)\citenamefont
  {Belavin}, \citenamefont {Polyakov}, \citenamefont {Schwartz},\ and\
  \citenamefont {Tyupkin}}]{Belavin:1975fg}%
  \BibitemOpen
  \bibfield  {author} {\bibinfo {author} {\bibfnamefont {A.}~\bibnamefont
  {Belavin}}, \bibinfo {author} {\bibfnamefont {A.~M.}\ \bibnamefont
  {Polyakov}}, \bibinfo {author} {\bibfnamefont {A.}~\bibnamefont {Schwartz}},
  \ and\ \bibinfo {author} {\bibfnamefont {Y.}~\bibnamefont {Tyupkin}},\ }\href
  {\doibase 10.1016/0370-2693(75)90163-X} {\bibfield  {journal} {\bibinfo
  {journal} {Phys.Lett.}\ }\textbf {\bibinfo {volume} {B59}},\ \bibinfo {pages}
  {85} (\bibinfo {year} {1975})}\BibitemShut {NoStop}%
\bibitem [{\citenamefont {Pham}(1983)}]{Pham1983}%
  \BibitemOpen
  \bibfield  {author} {\bibinfo {author} {\bibfnamefont {F.}~\bibnamefont
  {Pham}},\ }\href@noop {} {\bibfield  {journal} {\bibinfo  {journal} {Proc.
  Symp. Pure Math.}\ }\textbf {\bibinfo {volume} {2}},\ \bibinfo {pages} {319}
  (\bibinfo {year} {1983})}\BibitemShut {NoStop}%
\bibitem [{\citenamefont {Witten}(2010{\natexlab{a}})}]{Witten:2010cx}%
  \BibitemOpen
  \bibfield  {author} {\bibinfo {author} {\bibfnamefont {E.}~\bibnamefont
  {Witten}},\ }\href@noop {} {\bibfield  {journal} {\bibinfo  {journal}
  {preprint}\ ,\ \bibinfo {pages} {347}} (\bibinfo {year}
  {2010}{\natexlab{a}})},\ \Eprint {http://arxiv.org/abs/1001.2933}
  {arXiv:1001.2933 [hep-th]} \BibitemShut {NoStop}%
\bibitem [{\citenamefont {Witten}(2010{\natexlab{b}})}]{Witten:2010zr}%
  \BibitemOpen
  \bibfield  {author} {\bibinfo {author} {\bibfnamefont {E.}~\bibnamefont
  {Witten}},\ }\href@noop {} {\bibfield  {journal} {\bibinfo  {journal}
  {preprint}\ } (\bibinfo {year} {2010}{\natexlab{b}})},\ \Eprint
  {http://arxiv.org/abs/1009.6032} {arXiv:1009.6032 [hep-th]} \BibitemShut
  {NoStop}%
\bibitem [{\citenamefont {Tanizaki}\ and\ \citenamefont
  {Koike}(2014)}]{Tanizaki:2014xba}%
  \BibitemOpen
  \bibfield  {author} {\bibinfo {author} {\bibfnamefont {Y.}~\bibnamefont
  {Tanizaki}}\ and\ \bibinfo {author} {\bibfnamefont {T.}~\bibnamefont
  {Koike}},\ }\href@noop {} {\  (\bibinfo {year} {2014})},\ \Eprint
  {http://arxiv.org/abs/1406.2386} {arXiv:1406.2386 [math-ph]} \BibitemShut
  {NoStop}%
\bibitem [{\citenamefont {Bogomolny}(1980)}]{Bogomolny:1980ur}%
  \BibitemOpen
  \bibfield  {author} {\bibinfo {author} {\bibfnamefont {E.}~\bibnamefont
  {Bogomolny}},\ }\href {\doibase 10.1016/0370-2693(80)91014-X} {\bibfield
  {journal} {\bibinfo  {journal} {Phys.Lett.}\ }\textbf {\bibinfo {volume}
  {B91}},\ \bibinfo {pages} {431} (\bibinfo {year} {1980})}\BibitemShut
  {NoStop}%
\bibitem [{\citenamefont {Zinn-Justin}(1981)}]{ZinnJustin:1981dx}%
  \BibitemOpen
  \bibfield  {author} {\bibinfo {author} {\bibfnamefont {J.}~\bibnamefont
  {Zinn-Justin}},\ }\href {\doibase 10.1016/0550-3213(81)90197-8} {\bibfield
  {journal} {\bibinfo  {journal} {Nucl.Phys.}\ }\textbf {\bibinfo {volume}
  {B192}},\ \bibinfo {pages} {125} (\bibinfo {year} {1981})}\BibitemShut
  {NoStop}%
\bibitem [{\citenamefont {Dunne}\ and\ \citenamefont
  {Unsal}(2014)}]{Dunne:2014bca}%
  \BibitemOpen
  \bibfield  {author} {\bibinfo {author} {\bibfnamefont {G.~V.}\ \bibnamefont
  {Dunne}}\ and\ \bibinfo {author} {\bibfnamefont {M.}~\bibnamefont {Unsal}},\
  }\href@noop {} {\bibfield  {journal} {\bibinfo  {journal} {preprint}\ }
  (\bibinfo {year} {2014})},\ \Eprint {http://arxiv.org/abs/1401.5202}
  {arXiv:1401.5202 [hep-th]} \BibitemShut {NoStop}%
\bibitem [{\citenamefont {Argyres}\ and\ \citenamefont
  {Unsal}(2012{\natexlab{a}})}]{Argyres:2012vv}%
  \BibitemOpen
  \bibfield  {author} {\bibinfo {author} {\bibfnamefont {P.}~\bibnamefont
  {Argyres}}\ and\ \bibinfo {author} {\bibfnamefont {M.}~\bibnamefont
  {Unsal}},\ }\href {\doibase 10.1103/PhysRevLett.109.121601} {\bibfield
  {journal} {\bibinfo  {journal} {Phys.Rev.Lett.}\ }\textbf {\bibinfo {volume}
  {109}},\ \bibinfo {pages} {121601} (\bibinfo {year} {2012}{\natexlab{a}})},\
  \Eprint {http://arxiv.org/abs/1204.1661} {arXiv:1204.1661 [hep-th]}
  \BibitemShut {NoStop}%
\bibitem [{\citenamefont {Argyres}\ and\ \citenamefont
  {Unsal}(2012{\natexlab{b}})}]{Argyres:2012ka}%
  \BibitemOpen
  \bibfield  {author} {\bibinfo {author} {\bibfnamefont {P.~C.}\ \bibnamefont
  {Argyres}}\ and\ \bibinfo {author} {\bibfnamefont {M.}~\bibnamefont
  {Unsal}},\ }\href {\doibase 10.1007/JHEP08(2012)063} {\bibfield  {journal}
  {\bibinfo  {journal} {JHEP}\ }\textbf {\bibinfo {volume} {1208}},\ \bibinfo
  {pages} {063} (\bibinfo {year} {2012}{\natexlab{b}})},\ \Eprint
  {http://arxiv.org/abs/1206.1890} {arXiv:1206.1890 [hep-th]} \BibitemShut
  {NoStop}%
\bibitem [{\citenamefont {Dunne}\ and\ \citenamefont
  {Unsal}(2012)}]{Dunne:2012ae}%
  \BibitemOpen
  \bibfield  {author} {\bibinfo {author} {\bibfnamefont {G.~V.}\ \bibnamefont
  {Dunne}}\ and\ \bibinfo {author} {\bibfnamefont {M.}~\bibnamefont {Unsal}},\
  }\href {\doibase 10.1007/JHEP11(2012)170} {\bibfield  {journal} {\bibinfo
  {journal} {JHEP}\ }\textbf {\bibinfo {volume} {1211}},\ \bibinfo {pages}
  {170} (\bibinfo {year} {2012})},\ \Eprint {http://arxiv.org/abs/1210.2423}
  {arXiv:1210.2423 [hep-th]} \BibitemShut {NoStop}%
\bibitem [{\citenamefont {Dunne}\ and\ \citenamefont
  {Unsal}(2013)}]{Dunne:2012zk}%
  \BibitemOpen
  \bibfield  {author} {\bibinfo {author} {\bibfnamefont {G.~V.}\ \bibnamefont
  {Dunne}}\ and\ \bibinfo {author} {\bibfnamefont {M.}~\bibnamefont {Unsal}},\
  }\href {\doibase 10.1103/PhysRevD.87.025015} {\bibfield  {journal} {\bibinfo
  {journal} {Phys.Rev.}\ }\textbf {\bibinfo {volume} {D87}},\ \bibinfo {pages}
  {025015} (\bibinfo {year} {2013})},\ \Eprint {http://arxiv.org/abs/1210.3646}
  {arXiv:1210.3646 [hep-th]} \BibitemShut {NoStop}%
\bibitem [{\citenamefont {Cherman}\ \emph
  {et~al.}(2014{\natexlab{a}})\citenamefont {Cherman}, \citenamefont
  {Dorigoni}, \citenamefont {Dunne},\ and\ \citenamefont
  {Unsal}}]{Cherman:2013yfa}%
  \BibitemOpen
  \bibfield  {author} {\bibinfo {author} {\bibfnamefont {A.}~\bibnamefont
  {Cherman}}, \bibinfo {author} {\bibfnamefont {D.}~\bibnamefont {Dorigoni}},
  \bibinfo {author} {\bibfnamefont {G.~V.}\ \bibnamefont {Dunne}}, \ and\
  \bibinfo {author} {\bibfnamefont {M.}~\bibnamefont {Unsal}},\ }\href
  {\doibase 10.1103/PhysRevLett.112.021601} {\bibfield  {journal} {\bibinfo
  {journal} {Phys.Rev.Lett.}\ }\textbf {\bibinfo {volume} {112}},\ \bibinfo
  {pages} {021601} (\bibinfo {year} {2014}{\natexlab{a}})},\ \Eprint
  {http://arxiv.org/abs/1308.0127} {arXiv:1308.0127 [hep-th]} \BibitemShut
  {NoStop}%
\bibitem [{\citenamefont {Basar}\ \emph {et~al.}(2013)\citenamefont {Basar},
  \citenamefont {Dunne},\ and\ \citenamefont {Unsal}}]{Basar:2013eka}%
  \BibitemOpen
  \bibfield  {author} {\bibinfo {author} {\bibfnamefont {G.}~\bibnamefont
  {Basar}}, \bibinfo {author} {\bibfnamefont {G.~V.}\ \bibnamefont {Dunne}}, \
  and\ \bibinfo {author} {\bibfnamefont {M.}~\bibnamefont {Unsal}},\ }\href
  {\doibase 10.1007/JHEP10(2013)041} {\bibfield  {journal} {\bibinfo  {journal}
  {JHEP}\ }\textbf {\bibinfo {volume} {1310}},\ \bibinfo {pages} {041}
  (\bibinfo {year} {2013})},\ \Eprint {http://arxiv.org/abs/1308.1108}
  {arXiv:1308.1108 [hep-th]} \BibitemShut {NoStop}%
\bibitem [{\citenamefont {Cherman}\ \emph
  {et~al.}(2014{\natexlab{b}})\citenamefont {Cherman}, \citenamefont
  {Dorigoni},\ and\ \citenamefont {Unsal}}]{Cherman:2014ofa}%
  \BibitemOpen
  \bibfield  {author} {\bibinfo {author} {\bibfnamefont {A.}~\bibnamefont
  {Cherman}}, \bibinfo {author} {\bibfnamefont {D.}~\bibnamefont {Dorigoni}}, \
  and\ \bibinfo {author} {\bibfnamefont {M.}~\bibnamefont {Unsal}},\
  }\href@noop {} {\  (\bibinfo {year} {2014}{\natexlab{b}})},\ \Eprint
  {http://arxiv.org/abs/1403.1277} {arXiv:1403.1277 [hep-th]} \BibitemShut
  {NoStop}%
\bibitem [{\citenamefont {Misumi}\ \emph {et~al.}(2014)\citenamefont {Misumi},
  \citenamefont {Nitta},\ and\ \citenamefont {Sakai}}]{Misumi:2014jua}%
  \BibitemOpen
  \bibfield  {author} {\bibinfo {author} {\bibfnamefont {T.}~\bibnamefont
  {Misumi}}, \bibinfo {author} {\bibfnamefont {M.}~\bibnamefont {Nitta}}, \
  and\ \bibinfo {author} {\bibfnamefont {N.}~\bibnamefont {Sakai}},\ }\href
  {\doibase 10.1007/JHEP06(2014)164} {\bibfield  {journal} {\bibinfo  {journal}
  {JHEP}\ }\textbf {\bibinfo {volume} {1406}},\ \bibinfo {pages} {164}
  (\bibinfo {year} {2014})},\ \Eprint {http://arxiv.org/abs/1404.7225}
  {arXiv:1404.7225 [hep-th]} \BibitemShut {NoStop}%
\bibitem [{\citenamefont {Misumi}\ and\ \citenamefont
  {Kanazawa}(2014)}]{Misumi:2014raa}%
  \BibitemOpen
  \bibfield  {author} {\bibinfo {author} {\bibfnamefont {T.}~\bibnamefont
  {Misumi}}\ and\ \bibinfo {author} {\bibfnamefont {T.}~\bibnamefont
  {Kanazawa}},\ }\href {\doibase 10.1007/JHEP06(2014)181} {\bibfield  {journal}
  {\bibinfo  {journal} {JHEP}\ }\textbf {\bibinfo {volume} {1406}},\ \bibinfo
  {pages} {181} (\bibinfo {year} {2014})},\ \Eprint
  {http://arxiv.org/abs/1405.3113} {arXiv:1405.3113 [hep-ph]} \BibitemShut
  {NoStop}%
\bibitem [{Note2()}]{Note2}%
  \BibitemOpen
  \bibinfo {note} {There are also related developments in string theory, see
  {\protect \it e.g.} \cite
  {Marino:2007te,*Marino:2008ya,*Marino:2008vx,*Aniceto:2011nu,*Schiappa:2013opa}.}\BibitemShut
  {Stop}%
\bibitem [{Note3()}]{Note3}%
  \BibitemOpen
  \bibinfo {note} {The instanton and anti-instanton get interchanged if $-\pi
  /2 \le \alpha < 0$.}\BibitemShut {Stop}%
\bibitem [{\citenamefont {Harlow}\ \emph {et~al.}(2011)\citenamefont {Harlow},
  \citenamefont {Maltz},\ and\ \citenamefont {Witten}}]{Harlow:2011ny}%
  \BibitemOpen
  \bibfield  {author} {\bibinfo {author} {\bibfnamefont {D.}~\bibnamefont
  {Harlow}}, \bibinfo {author} {\bibfnamefont {J.}~\bibnamefont {Maltz}}, \
  and\ \bibinfo {author} {\bibfnamefont {E.}~\bibnamefont {Witten}},\ }\href
  {\doibase 10.1007/JHEP12(2011)071} {\bibfield  {journal} {\bibinfo  {journal}
  {JHEP}\ }\textbf {\bibinfo {volume} {1112}},\ \bibinfo {pages} {071}
  (\bibinfo {year} {2011})},\ \Eprint {http://arxiv.org/abs/1108.4417}
  {arXiv:1108.4417 [hep-th]} \BibitemShut {NoStop}%
\bibitem [{Note4()}]{Note4}%
  \BibitemOpen
  \bibinfo {note} {How to carry out this program completely for
  configuration-space path integrals is an open problem, but the present
  understanding is already sufficient for our Gaussian analysis.}\BibitemShut
  {Stop}%
\bibitem [{\citenamefont {Levkov}\ and\ \citenamefont
  {Sibiryakov}(2005)}]{Levkov:2004ij}%
  \BibitemOpen
  \bibfield  {author} {\bibinfo {author} {\bibfnamefont {D.}~\bibnamefont
  {Levkov}}\ and\ \bibinfo {author} {\bibfnamefont {S.}~\bibnamefont
  {Sibiryakov}},\ }\href {\doibase 10.1134/1.1887914} {\bibfield  {journal}
  {\bibinfo  {journal} {JETP Lett.}\ }\textbf {\bibinfo {volume} {81}},\
  \bibinfo {pages} {53} (\bibinfo {year} {2005})},\ \Eprint
  {http://arxiv.org/abs/hep-th/0412253} {arXiv:hep-th/0412253 [hep-th]}
  \BibitemShut {NoStop}%
\bibitem [{\citenamefont {Marino}\ \emph {et~al.}(2008)\citenamefont {Marino},
  \citenamefont {Schiappa},\ and\ \citenamefont {Weiss}}]{Marino:2007te}%
  \BibitemOpen
  \bibfield  {author} {\bibinfo {author} {\bibfnamefont {M.}~\bibnamefont
  {Marino}}, \bibinfo {author} {\bibfnamefont {R.}~\bibnamefont {Schiappa}}, \
  and\ \bibinfo {author} {\bibfnamefont {M.}~\bibnamefont {Weiss}},\
  }\href@noop {} {\bibfield  {journal} {\bibinfo  {journal}
  {Commun.Num.Theor.Phys.}\ }\textbf {\bibinfo {volume} {2}},\ \bibinfo {pages}
  {349} (\bibinfo {year} {2008})},\ \Eprint {http://arxiv.org/abs/0711.1954}
  {arXiv:0711.1954 [hep-th]} \BibitemShut {NoStop}%
\bibitem [{\citenamefont {Marino}(2008)}]{Marino:2008ya}%
  \BibitemOpen
  \bibfield  {author} {\bibinfo {author} {\bibfnamefont {M.}~\bibnamefont
  {Marino}},\ }\href {\doibase 10.1088/1126-6708/2008/12/114} {\bibfield
  {journal} {\bibinfo  {journal} {JHEP}\ }\textbf {\bibinfo {volume} {0812}},\
  \bibinfo {pages} {114} (\bibinfo {year} {2008})},\ \Eprint
  {http://arxiv.org/abs/0805.3033} {arXiv:0805.3033 [hep-th]} \BibitemShut
  {NoStop}%
\bibitem [{\citenamefont {Marino}\ \emph {et~al.}(2009)\citenamefont {Marino},
  \citenamefont {Schiappa},\ and\ \citenamefont {Weiss}}]{Marino:2008vx}%
  \BibitemOpen
  \bibfield  {author} {\bibinfo {author} {\bibfnamefont {M.}~\bibnamefont
  {Marino}}, \bibinfo {author} {\bibfnamefont {R.}~\bibnamefont {Schiappa}}, \
  and\ \bibinfo {author} {\bibfnamefont {M.}~\bibnamefont {Weiss}},\ }\href
  {\doibase 10.1063/1.3097755} {\bibfield  {journal} {\bibinfo  {journal}
  {J.Math.Phys.}\ }\textbf {\bibinfo {volume} {50}},\ \bibinfo {pages} {052301}
  (\bibinfo {year} {2009})},\ \Eprint {http://arxiv.org/abs/0809.2619}
  {arXiv:0809.2619 [hep-th]} \BibitemShut {NoStop}%
\bibitem [{\citenamefont {Aniceto}\ \emph {et~al.}(2012)\citenamefont
  {Aniceto}, \citenamefont {Schiappa},\ and\ \citenamefont
  {Vonk}}]{Aniceto:2011nu}%
  \BibitemOpen
  \bibfield  {author} {\bibinfo {author} {\bibfnamefont {I.}~\bibnamefont
  {Aniceto}}, \bibinfo {author} {\bibfnamefont {R.}~\bibnamefont {Schiappa}}, \
  and\ \bibinfo {author} {\bibfnamefont {M.}~\bibnamefont {Vonk}},\ }\href@noop
  {} {\bibfield  {journal} {\bibinfo  {journal} {Commun.Num.Theor.Phys.}\
  }\textbf {\bibinfo {volume} {6}},\ \bibinfo {pages} {339} (\bibinfo {year}
  {2012})},\ \Eprint {http://arxiv.org/abs/1106.5922} {arXiv:1106.5922
  [hep-th]} \BibitemShut {NoStop}%
\bibitem [{\citenamefont {Schiappa}\ and\ \citenamefont
  {Vaz}(2014)}]{Schiappa:2013opa}%
  \BibitemOpen
  \bibfield  {author} {\bibinfo {author} {\bibfnamefont {R.}~\bibnamefont
  {Schiappa}}\ and\ \bibinfo {author} {\bibfnamefont {R.}~\bibnamefont {Vaz}},\
  }\href {\doibase 10.1007/s00220-014-2028-7} {\bibfield  {journal} {\bibinfo
  {journal} {Commun.Math.Phys.}\ }\textbf {\bibinfo {volume} {330}},\ \bibinfo
  {pages} {655} (\bibinfo {year} {2014})},\ \Eprint
  {http://arxiv.org/abs/1302.5138} {arXiv:1302.5138 [hep-th]} \BibitemShut
  {NoStop}%
\end{thebibliography}%
\end{document}